\documentstyle[prl,multicol,aps]{revtex}
\newcommand{\BEQ}{\begin{equation}}
\newcommand{\EEQ}{\end{equation}}
\newcommand{\BEA}{\begin{eqnarray}}
\newcommand{\EEA}{\end{eqnarray}}

\renewcommand{\d}{{\rm d}}
\renewcommand{\top }{ t^{\prime } }

\newcommand{\x}{\hat{x}}
\newcommand{\p}{\hat{p}}

\newcommand{\e}{\hat{\eta}}
\newcommand{\iH}{\hat{H}}
\newcommand{\PP}{{\cal P}}
\newcommand{\MM}{{\cal M}}
\newcommand{\rh}{\hat{\rho}}

\begin{document}
\draft
\title
{Mean-field theory of quantum brownian motion}

\author{A.E. Allahverdyan$^{1,3)}$, R. Balian$^{1)}$ 
and Th.M. Nieuwenhuizen$^{2)}$}
\address{$^{1)}$ CEA/Saclay, Service de Physique Theorique, 
F-91191 Gif-sur-Yvette Cedex, France, \\ 
$^{2)}$ Department of Physics and Astronomy,\\ 
University of Amsterdam,
Valckenierstraat 65, 1018 XE Amsterdam, The Netherlands 
\\ $^{3)}$Yerevan Physics Institute,
Alikhanian Brothers St. 2, Yerevan 375036, Armenia }
\maketitle

\begin{abstract}
We investigate a mean-field approach to a quantum brownian particle 
interacting with a quantum thermal bath at temperature $T$, 
and subjected to a non-linear potential.
An exact, partially classical description of quantum brownian 
motion is proposed, which uses negative
probabilities in its intermediate steps.
It is shown that properties of the quantum particle can be mapped to those 
of two classical brownian particles in a common potential, where
one of them interacts with the quantum bath, 
whereas another one interacts with a classical bath at zero
temperature. Due to damping the system allows
a unique and non-singular classical limit at $\hbar \to 0$. For 
high $T$ the stationary state becomes explicitly classical.
The low-temperature case is studied through an effective Fokker-Planck
equation. Non-trivial purely quantum correlation effects
between the two particles are found. 
\end{abstract}

\section{Introduction}
The main conceptual problem in quantum mechanics remains the link
between the quantum and the classical worlds. Therefore,
significant efforts were made over years to understand at least  
part of the quantum world in classical notions. Curiously, the 
quasiclassical domain, which should be the main subject of this 
understanding, is still itself under extensive investigation. 
Indeed, it is known to be non-trivial; in a sense it can be
even more complex that the classical and quantum extremes alone.
It is important to realize in this context that the quasiclassical domain 
is not exhausted by the conventional ansatz $\hbar \to 0$
\cite{ballentine,berry,tatarinov}, since 
this limit is singular (therefore some a priori concepts 
similar to coarse-graining are sometimes involved \cite{tatarinov}), 
and since it does not commute 
with other limits of physical interest, e.g. the limit of large times.

One of the established approaches to the quasiclassical domain 
is a collection of mean-field methods known as Gaussian decoupling 
procedure or time-dependent variational approximation 
\cite{heller,balian,shavo,klaud}. 
Mean-field (variational, Hartree-Fock) methods
are well-known in the quantum theory, 
and were applied for a while
in many different areas. This set of methods appeared 
to be especially suitable for the quasiclassical domain, since it attempts
to realize in a simple and straightforward 
way the above-mentioned program of understanding
the quantum theory in classical terms. An impressive amount of
experimental
confirmations, in particular in quantum chemistry and atomic physics 
\cite{heller},
numerical and self-consistency checks were made for those methods.
Therefore,
they have already become a well-formulated and sound approach.

In the present paper we apply this mean-field method to 
the simplest quantum dissipative system: A quantum brownian particle
interacting with a thermal bath. There are
specific reasons to study this type of quantum systems in the context of
the
above-mentioned problems. 
In contrast to closed Hamiltonian systems, quantum
dissipative systems are more reliable candidates to understand their
physics
in classical terms. 
Indeed, a non-unitary evolution provides a natural mechanism of
decoherence
\cite{zurek}, and the need of artificial coarse-graining procedures is
gotten 
rid of. In the light of this conceptual advantage
it should be surprising that 
the basic understanding of their theory 
is still rather fragmentary. This is so mainly
because one can use neither general properties of 
unitary evolution, which describes closed systems, 
nor markovian properties of the classical stochastic dynamics, since
due to the correlation time $\hbar/T$, which is relevant at low
temperatures, the corresponding statistical dynamics is essentially
non-markovian. 
Both these dynamical properties provide important general information,
which is, thus, absent in the quantum dissipative case \cite{weiss}.
The only exceptions are weakly 
damped high-temperature systems, where a copulation 
of those two things is possible, namely, an influence of the bath is 
described classically, whereas the rest remains quantum-mechanical 
\cite{zurek}. We will not be concerned with them within the present paper.
A fairly general theory of strongly-damped and/or low-temperature
dissipative systems is still under construction, though some suggestive
results were obtained recently \cite{AN},
where for a class of systems two of us proposed a consistent 
statistical thermodynamical theory of quantum brownian motion.

Our plan is the following. In section II we will first briefly recall 
some known facts on quantum Langevin equations and quantum noise. 
Here we obtain also generalized Wigner-Moyal and von Neumann equations, 
which are exact consequences of the quantum dynamics, but describe 
the brownian particle in almost classical terms.
The Gaussian approximation will be presented in section III, 
where we will obtain the basic equations of the present paper, 
and draw some general conclusions. Here we will discuss, in particular, 
what is the constructive role of friction when establishing
quantum-classical 
transition. The effective Fokker-Planck dynamics will be discussed in 
section IV. In section V we study low-temperature properties of the model.
We conclude in the last section.

\section{Quantum Langevin equation and Liouville-Moyal equation}

\subsection{Quantum Langevin equation}

This fundamental equation of quantum brownian motion theory
is derived from the exact hamiltonian description of a subsystem
(brownian particle)
and a thermal bath, when tracing out the degrees of freedom of the bath.
The standard assumption is 
that at the moment $t=0$ the states of the
subsystem and the bath were decoupled from each other,   
and the bath was in equilibrium at temperature $T$
\cite{weiss,gardiner,AN}. Further, the influence of the particle 
to the bath is assumed to be sufficiently small; thus, only
the linear modes of the bath are excited, and the interaction 
between the particle and the bath is linear. Since the dynamics of 
the bath is linear, it can be solved exactly. Following this 
line of exact calculations \cite{weiss}, one derives the quantum 
Langevin equation 
\BEA
\label{01}
&&\dot{\x }=\frac{\p }{m},
\nonumber\\
&&\dot{\p}+\frac{1}{m}\int _0^t\d t^{\prime}
\gamma (t-t')\p (t^{\prime}) +V^{\prime}(\x )
=-\gamma\Gamma e^{-\Gamma t}\x (0)+\e (t),
\EEA
where $\p (t)$ $\x (t)$ are Heisenberg operators of momentum and
coordinate,
and $V(x)$ is an external potential. The parameter
$\gamma $ is the damping constant, which determines the interaction
between the
bath and the particle. 
For $\gamma\to 0 $ one gets from Eq.~(\ref{01}) the usual Heisenberg
equations.
$\Gamma$ is the maximal characteristic frequency of the bath, and it
determines the
retardation time of the friction 
kernel 
\BEA\label{kernel}
\gamma (t)=\gamma \Gamma e^{-\Gamma |t|}.
\EEA
The operator $\e (t)$ is the random noise, which appeared due to the
uncertain character of the initial (equilibrium) distribution
of the bath. This noise can be shown to be gaussian, due to the fact that
the thermal bath is a harmonic system and was in equilibrium.
It has the following properties:
\begin{equation}
\label{02}
K(t)=\frac{1}{2}\langle \e (t) \e (0) +  
\e (0) \e (t) \rangle _{\e } \equiv 
\frac{1}{2}\langle \e (t); \e (0)\rangle _{\e }=
\frac{\hbar \gamma }{\pi}\int_0^{\infty}\d \omega 
\,{\omega \coth \left (\frac{\hbar\omega\beta}{2}\right)
\cos (\omega t)} \frac{1}{1+(\omega /\Gamma )^2},
\end{equation}
\begin{equation}
\label{022}
\e (t) \e (0) -  
\e (0) \e (t) \equiv [\e (t), \e (0)]=
i\hbar \frac{\partial \gamma (t)}{\partial t}.
\end{equation}
Hereafter we ase
\BEA
\langle \hat{A}; \hat{B}\rangle \equiv\frac{1}{2} \langle \hat{A} \hat{B}
+ \hat{B}\hat{A}\rangle ,\qquad
[\hat{A},\hat{B}]\equiv \hat{A}\hat{B}-\hat{B}\hat{A},\qquad
[\hat{A},\hat{B}]_{+}\equiv \hat{A}\hat{B}+\hat{B}\hat{A}
\EEA
for any operators $\hat{A}$, $\hat{B}$.

The connection between properties of the noise and the friction kernel
is the consequence of quantum fluctuation-dissipation theorem 
\cite{weiss}. 
Eq.~(\ref{01}) with physically suitable forms of the potential
and friction describes a rich 
variety of physical phenomena 
(see references in \cite{weiss,gardiner,AN}). 

\subsubsection{Quasi-Ohmic limit}

In the present paper 
we shall restrict ourselves to the {\it quasi-Ohmic} case, where $\Gamma$
is much larger than other characteristic times, but still finite. 
The main reason of this approximation is to have an exact 
equation for the Wigner function of the brownian particle, 
which will be derived in the next subsection.

For the quantum noise one has that $K(t)=-\ln (\Gamma t)>0$ for 
small times, and for $t\gg 1/\Gamma$, $K(t)$ is anticorrelated with the 
universal correlation time $\hbar\beta /2\pi$: 
\begin{equation}
\label{5}
K(t)=-\frac{\pi \gamma T^2}{\hbar}\left [
\sinh \left (\frac{\pi t}{\beta\hbar}\right )
\right ]^{-2},
\end{equation}
Being coherent, the low-temperature quantum thermal bath 
neccesarily generates a colored noise.
The classical white noise situation is recovered when taking the 
high-temperature limit ($\hbar\beta\to 0$). In general, this should 
be done before the limit $\Gamma\to\infty$. Notice that in contrast
with the classical case, the quantum noise does not disappear for
$T\to 0$, since even in this limit the initial state of the quantum
thermal bath remains indeterminate.

In the quasi-Ohmic regime one can expand the memory kernel 
for the friction in Eq.~(\ref{01})
\begin{equation}
\label{1.1}
\int _0^t\d t^{\prime}
e^{-\Gamma (t-t^{\prime})}\p (t^{\prime})=\frac{1}{\Gamma}\p (t)
-\frac{1}{\Gamma ^2}\dot{\p}(t)-\frac{1}{\Gamma }e^{-\Gamma t}\p (0)
+\frac{1}{\Gamma ^2}e^{-\Gamma t}\dot{\p}(0).
\end{equation}
For $t>0$ and large $\Gamma $ the exponentially small factors
depending on the initial conditions
can be omitted in Eqs.~(\ref{01}, \ref{1.1}), and we finally get
\begin{equation}
\label{1.2}
\dot{\p}+\frac{\gamma}{m}\p (t)+ V^{\prime}(\x )=\e (t),
\end{equation}
In other words, for not very small times one can regularize only the
noise,
but keep the friction local, when considering the quasi Ohmic case. 

Let us finally notice the following general comutation relation
between the noise and an arbitrary operator $\hat{A}(t)$
of the particle \cite{sen,weiss}:
\BEA
\label{samarkand}
[\hat{A}(t),\hat{\eta }(s)] = \int _0^t \d t'
[\hat{A}(t),\hat{x }(t')]\frac{\d }{\d s}\gamma (s-t')
\EEA
Here the coordinate operator appeared just because the interaction 
with the thermal bath is taking place through the coordinate
\cite{sen,weiss}.
In the quasi-Ohmic case this will go to
\BEA
\label{tebriz}
[\hat{A}(t),\hat{\eta }(s)] = 2\gamma \frac{\d }{\d s}\{\theta (t-s)
[\hat{A}(t),\hat{x }(s)]\},
\EEA
where $\theta (t-s)$ is defined to be $1/2$ at $t=s$. 
It is seen that $[\e (t),\x (t)]=[\e (t),\p (t)]=0$. One could
attribute these relations to causality, but they are an emergent
property of the quasi-Ohmic limit.

\subsection{Generalized Wigner-Moyal equation}
The quantum Langevin equation (\ref{1.2}) is a non-linear operator
equation, and as such it can hardly be handled directly. Here we will
present a generalized Wigner-Moyal equation for the corresponding Wigner
function of the particle, which exactly corresponds to Eq.~(\ref{1.2})
in the same sense as the pure Heisenberg equations correspond to the
usual Wigner-Moyal equation. Besides technical advantages
which will be used further, this equation presents an 
interesting account for an exact description of the 
non-linear quantum problem through proper classical 
terms.

One is looking for an equation for the Wigner function:
\BEA
\label{kismet}
&&W(x,p,t)= \langle {\rm tr}\rho _0~\hat{W}(\hat{x},\hat{p},t)
\rangle _{\hat{\eta}},
\nonumber \\
&& \hat{W}(\hat{x},\hat{p},t)=\int \frac{\d a~\d b}{4\pi ^2}
\exp (-iax-ibp+ia\hat{x}(t)+ib\hat{p}(t)),
\EEA
where 
$\rho _0$ is the initial (at the moment $t=0$) 
density matrix of the brownian particle.
$\hat{W}(\hat{x},\hat{p},t)$ is the support of the Wigner function, 
which can be viewed as a quantum analogue of the delta-function.
In the classical limit, where $\hat{x}$ and $\hat{p}$ approxmately
commute, $W(x,p,t)$ tends to the ordinary probability distribution of
the coordinate and momentum.

Since the equation for $W(x,p)$ is expected to be linear, 
one is interested by an equation for $\hat{W}(\hat{x},\hat{p})$, whereas
the averages can be taken later. The derivation of this equation is fairly
straightforward, since it uses Eqs.~(\ref{1.2}, \ref{kismet}) and the 
standard commutation relation $[\x (t),\p (t)]=i\hbar $. We will write
only the final result:
\BEA
\label{nizami}
\frac{\partial \hat{W}(\hat{x},\hat{p},t)}{\partial t}=
-\frac{\partial }{\partial x}(\frac{p}{m}\hat{W})+
\frac{\partial }{\partial p}([V'(x)+\frac{\gamma}{m}p]\hat{W})
+\sum_{n=1}^{\infty}\frac{(i\hbar/2)^{2n}}{(2n+1)!}~
\frac{\partial ^{2n+1}V}{\partial x^{2n+1}}~
\frac{\partial ^{2n+1}\hat{W}}{\partial p^{2n+1}}
-\frac{\partial }{\partial p}(\hat{\eta}\hat{W}).
\EEA
The first three terms in the r.h.s of this equation are the standard drift 
terms of the classical Liouville equation, and the sum represents a purely 
quantum correction, which comes from the non-linearity of the potential.

Our object of interest is the last term, which upon averaging will look
like
\BEA
\label{khorezm}
\frac{\partial }{\partial p}\langle \hat{\eta}(t)\hat{W}(t)
\rangle _{\hat{\eta}}.
\EEA
Notice that $\hat{\eta}(t)$, $\hat{W}(t)$ commute, since 
$\hat{\eta}(t)$ commutes with $x(t)$, $p(t)$, and $\hat{W}(t)$
can be presented as a product of two terms, which depend only on
$\x (t)$ and $\p (t)$ correspondingly. 

Let us adopt the following formal expansion for 
$\hat{W}(t)$ 

\BEA
\label{astara1}
&&\hat{W}(t|\eta )= \hat{W}(t|0 )+\frac{1}{n!}\sum_{n=1}^{\infty}
\int _0^t \d s_1...\d s_n R(s_1,..,s_n)
\Pi [\hat{\eta} (s_1)...\hat{\eta } (s_n)],
\\
\label{astara2}
&& \Pi [\hat{\eta} (s_1)...\hat{\eta } (s_n)]=
\frac{1}{n!}\sum_{i_1\not = ...\not = i_n}
\hat{\eta} (s_{i_1})...\hat{\eta } (s_{i_n}).
\EEA
Here $\Pi [\hat{\eta} (s_1)...\hat{\eta } (s_n)]$ is the symmetrized
product; the coefficients $R(s_1,..,s_n)$ are c-numbers.
Due to the fact that $[\hat{W}(t),\e (t)]=0$ one has:
\BEA
\label{kazan}
\langle \hat{\eta}(t)\hat{W}(t)
\rangle _{\hat{\eta}}=\langle \hat{\eta}(t);\hat{W}(t)
\rangle _{\hat{\eta}}=
\sum_{n=1}^{\infty}\frac{1}{n!}
\int _0^{t} \d s_1...\d s_n R(s_1,..,s_n)\langle
\hat{\eta}(t);\Pi [\hat{\eta} (s_1)...\hat{\eta } (s_n)]
\rangle _{\hat{\eta}}.
\EEA
Since $\e (t)$ is a gaussian random operator with
$\langle\e (t)\rangle _{\hat{\eta}}=0$, one can 
use Wick's theorem:
The correlation of an odd number of $\e$'s vanishes. The correlation
of an even number of $\e$'s is equal to the sum of products of pair 
correlations, the sum being taken over all pairings. 
For example:
\BEA
\langle \e (t_1)\e (t_2)\e (t_3)\e (t_4)\rangle _{\e}=
\langle \e (t_1)\e (t_2)\rangle _{\e}
\langle \e (t_3)\e (t_4)\rangle _{\e}+
\langle \e (t_1)\e (t_3)\rangle _{\e}
\langle \e (t_2)\e (t_4)\rangle _{\e}+
\langle \e (t_1)\e (t_4)\rangle _{\e}
\langle \e (t_2)\e (t_3)\rangle _{\e}
\EEA
In this way one derives: 
\BEA
\label{dub}
\langle
\hat{\eta}(t);\Pi [\hat{\eta} (s_1)...\hat{\eta } (s_n)]
\rangle _{\hat{\eta}}=
\sum_{\alpha =1}^n\langle \hat{\eta}(t);
\hat{\eta} (s_{\alpha})      \rangle _{\hat{\eta}}~\langle
\Pi [\hat{\eta} (s_1)...\hat{\eta} (s_{\alpha -1})
\hat{\eta} (s_{\alpha +1})...\hat{\eta } (s_n)]\rangle _{\hat{\eta}}.
\EEA
Having substituted this equation to Eq.~(\ref{kazan}) one gets
\BEA
\label{kuba}
\langle \hat{\eta}(t)\hat{W}(t)
\rangle _{\hat{\eta}}=
\sum_{n=1}^{\infty} \frac{1}{(n-1)!}\int _0^{t} \d s_1
\langle \hat{\eta}(t);\hat{\eta} (s_{1}) \rangle _{\hat{\eta}}
\int _0^{t} \d s_2...\d s_n R(s_1,..,s_n)\langle
\Pi [\hat{\eta} (s_2)...\hat{\eta } (s_n)]
\rangle _{\hat{\eta}}.
\EEA
Now one notices that the only feature of the quantum noise
which enters here is the autocorrelation function
$
\langle \hat{\eta}(t);\hat{\eta} (s) \rangle _{\hat{\eta}}=K(t-s)
$
because in the end $\langle
\Pi [\hat{\eta} (s_2)...\hat{\eta } (s_n)]
\rangle _{\hat{\eta}}$ can be expressed through it using Eq.~(\ref{dub})
several times. So nothing will change if we replace $\e (t)$ 
in Eq.~(\ref{kuba}) by
a {\it classical} gaussian noise $\eta (t)$ which has the same 
autocorrelation function $K(t-s)$,
\BEA
\langle \hat{\eta}(t)\hat{W}(t)
\rangle _{\hat{\eta}}
=\langle \eta(t)\hat{W}(t)
\rangle _{\eta}.
\EEA
Now we substitute this result into Eqs.~(\ref{nizami}), take the
average over the initial state, but take out the average over the
classical 
noise:
\BEQ
\label{moyal}
\partial _{t}w = -\partial _{x}(\frac{p}{m}w)
+\partial _{p}([V'(x)+\frac{\gamma}{m}p-\eta (t)]w)
+\sum_{n=1}^{\infty}\frac{(i\hbar/2)^{2n}}{(2n+1)!}~
\frac{\partial ^{2n+1}V}{\partial x^{2n+1}}~
\frac{\partial ^{2n+1}w}{\partial p^{2n+1}},
\EEQ
where the true Wigner function $W(x,p,t)$ will be obtained by
averaging over the classical noise $\eta (t)$:
\BEA
\label{abdurahman}
W(x,p,t)= \langle w(x,p,t)\rangle _{\eta},
\EEA
where $w(x,p,t)$ is an auxiliary object.
Thus in the quasi-Ohmic limit one can treat the quantum noise as a purely
classical object as far as the Wigner function is concerned. Notice that
this result is exact and relies only on the quasi-Ohmic limit.
In that respect it is different from the description through semiclassical
Langevin equations, where in the overdamped (large $\gamma$)
limit one also gets an analogue of Eq.~(\ref{moyal}) but without
the $\hbar$-dependent terms.

\subsection{Generalized von Neumann equation}
Here we will investigate the von Neumann equation for the density
matrix, which corresponds to the Wigner function $w(x,p)$:
\BEQ
\label{wiwi}
w(p,q) = \frac{1}{2\pi \hbar}\int \d u ~r(q-
\frac{u}{2}, q+\frac{u}{2})e^{ipu/\hbar}.
\EEQ

Since Eq.~(\ref{moyal}) is linear, one directly obtains for
$r(x,x',t)=\langle x'|\hat{r}(t)|x\rangle$
\BEQ
\label{korshun}
-i\hbar \frac{\partial}{\partial t}r(x,x',t)
=\left [ \frac{\hbar ^2}{2m}\frac{\partial ^2}{\partial x^2}- 
\frac{\hbar ^2}{2m}\frac{\partial ^2}{\partial x'^2} - V(x) + V(x') + \eta
(t)
(x-x') + \frac{i\gamma \hbar }{m}(x-x') (\frac{\partial }{\partial x} -
\frac{\partial }{\partial x'})
\right ] r(x,x',t),
\EEQ
or in the equivalent operator notations:
\BEQ
\label{iastreb2}
\frac{\d\hat{r} }{\d t}=
-\frac{i }{\hbar}[\iH -\x \eta,\hat{r}]+\frac{i\gamma }{2\hbar m}[\{
\p,\hat{r} 
 \}, \x ].
\EEQ
To obtain this equation from Eq.~(\ref{moyal}) in quick way, one can
use the following correspondence between operators $\hat{A}$, $\hat{B}$
and their
phase-space representations $A^{(c)}(x,p)$, $B^{(c)}(x,p)$:
\BEA
\frac{i}{\hbar}[\hat{A},\hat{B}]=\partial _pA^{(c)}\partial
_xB^{(c)}
-\partial _pB^{(c)}\partial _xA^{(c)},\qquad
[\hat{A},\hat{B}]_{+}=2A^{(c)}B^{(c)},
\EEA
and take into account that $x^{(c)}=x$, $p^{(c)}=p$, $r^{(c)}=w(x,p)$.

The true density matrix $\hat{\rho }$ can be obtained after averaging by 
the classical noise $\eta (t)$:
\BEQ
\label{korshun1}
\hat{\rho }(t)=\langle \hat{r}(t)\rangle _{\eta }.
\EEQ
Therefore, all possible averages are obtained as
\BEQ
\label{iastreb1}
{\rm tr}(\rh \hat{A})=\langle {\rm tr}(\hat{r}\hat{A})\rangle _{\eta },
\EEQ
where an operator $\hat{A}$ lives in the Hilbert space of the brownian 
particle. 

It should be stressed that $\hat{r}$ is not a density matrix itself,
but rather a tool to calculate averages. Indeed starting from 
Eq.~(\ref{iastreb2}) one easily gets
\BEQ
\label{iastreb3}
\frac{\d }{\d t}{\rm tr}\,\hat{r}
=0\, \to \, {\rm tr}\,\hat{r} (t) = 1,
\EEQ
\BEQ
\label{iastreb4}
\frac{\d }{\d t}{\rm tr}(\hat{r}(t)^2)
=\frac{\gamma }{m}{\rm tr}(\hat{r}(t)^2)\, \to \,
{\rm tr}(\hat{r}(t)^2)=e^{\gamma t/m}\,{\rm tr}(\hat{r}(0)^2).
\EEQ
These indicate that $r$ has negative eigenvalues for $t>0$, the absolute 
value of which grows with time. However, positive eigenvalues
compensate this growth in such a way that Eq.~(\ref{iastreb3}) holds.
In particular, Eqs.~(\ref{iastreb3}, \ref{iastreb4}) show that a
dissipative systems cannot be described by a wave function even if the
averaging over the noise is postponed.

Thus we see that the ``classicalization'' of the noise is not just a
technical procedure, but it has to be 
accompanied with a change of interpretation:
The unaveraged density matrix $\hat{r}$ is not a true density matrix, 
since it does have negative eigenvalues. 
In other words, explicitly classical components of the dynamics lead to 
the appearance of negative probabilities.
On the other hand, there are no
reasons to consider those negative eigenvalues as something unphysical:
Our derivation of Eqs.~(\ref{moyal}, \ref{iastreb2}) was exact, and
later we will present other indications that neither the true density
matrix, nor averages calculated according to Eq.~(\ref{iastreb1}) show
unphysical properties.
The situation is the same as for the Wigner function at a given time,
which is not a positive probability density, but can be safely used  
to evaluate expectation values by integration.

The appearance of negative probabilities as a result of imposing
partially classical properties has certain analogies with the
classical interoperation of quantum entanglement, and in particular
Einstein-Podolsky-Rosen phenomenon. This interpretation also uses
negative (though not directly observable) probabilities \cite{muck}.

Finally we would like to mention that the reported non-positive
character of the unaveraged density matrix has nothing to do with the
known technical problem which also appears through non-positive
(averaged) density matrices in certain quantum markovian diffusion
equations \cite{haake}. There the problem is merely technical and
arises due to the fact that the markovian approximation in the quantum
theory of open systems is essentially time-inhomogeneous, so that its
careless use leads to such  problems. Indeed, the problem disappears after
a
more consistent treatment of the situation \cite{haake}. In our case,
in contrast, the non-positive character of the unaveraged density
matrix is an exact consequence of our attempt to handle the quantum
noise classically.

\section{Gaussian decoupling procedure}

Eq.~(\ref{moyal}) is, of course, intractable in general. It joins all
technical difficulties of the classical Liuoville equation with a given
noise and friction and those of the pure Moyal equation. In other words
some 
substantial simplifications are necessary to proceed further. Here we will 
apply the mean-field approach, namely, a solution of Eq.~(\ref{moyal})
will
be looked through its moments, and the gaussian decoupling procedure 
will be applied
to the higher-order moments. 
By its spirit this is very similar to the Grad method in the kinetic
theory of rarefied gases \cite{grad}, and has been applied recently
in quantum theory of closed systems as well \cite{shavo,heller}.
It is clear that the consistency of this
approximation should be checked together with the final solution. At the
moment we will notice only that since its application is connected with
the
weakness of quantum fluctuations, it will have a reliable range of
validity in 
the quasiclassical domain.

The working variables will be 
\begin{eqnarray}
\label{set}
&& d_{x}=\langle \x \rangle =\int \d p~\d x ~xw(p,x), \\
\label{setu1}
&&d_{p}=\langle \p \rangle =\int \d p~\d x ~pw(p,x), \\
\label{setu2}
&&d_{xx}=\langle (\x-\langle \x \rangle )^2\rangle 
  =\int \d p~\d x~(x-\langle x \rangle )^2w(p,x), \\ 
\label{setu3}
&&d_{pp}=\langle (\p-\langle \p \rangle )^2 \rangle 
  =\int \d p~\d x~(x-\langle x \rangle )^2w(p,x), \\ 
\label{setu4}
&&d_{xp}=\frac{1}{2}\langle (\x-\langle \x \rangle )(\p-\langle \p \rangle
)+
(\p-\langle \p \rangle )(\x-\langle \x \rangle )\rangle
=\int \d p~\d x ~(x-\langle x \rangle )(p-\langle p \rangle )w(p,x). 
\end{eqnarray}
For the higher-order correlations one assumes the gaussian decoupling:
\begin{eqnarray}
\label{set2}
&&\int \d p~\d x~w(p,x)(x-\langle x \rangle )^{2n} =
(2n-1)!!\left [
\int \d p~\d x~w(p,x)(x-\langle x \rangle )^{2}
\right ]^n \\
&&\int \d p~\d x~w(p,x)
(p-\langle p \rangle )(x-\langle x \rangle )^{2n+1}
=(2n+1)!!\left [\int \d p~\d x~w(p,x)(p-\langle p \rangle )
\right ]\left [
\int \d p~\d x~w(p,x)(x-\langle x \rangle )^{2}
\right ]^n.
\label{set3}
\end{eqnarray}
This just means that $w(p,x)$ is restricted to the subspace of
gaussian functions:
\BEA
&&w\left (x,p|d_{xx},d_{xp}, d_{pp}, d_{x}, d_{p}
\right )= \frac{1}{2\pi\sqrt{\Delta}}
\exp\left ( -\frac{1}{2\Delta}\left [
d_{pp}(x-d_x)^2+d_{xx}(p-d_p)^2-2d_{xp}(x-d_x)(p-d_p)
\right ]\right ),\nonumber\\
&& \Delta =d_{xx}d_{pp}-d_{xp}^2.
\EEA
Applying Eqs.~(\ref{set}-\ref{set3})
in Eq.~(\ref{moyal}) one gets the following equations of motions, which 
are now classical equations for classical variables:
\begin{eqnarray}
\label{nippon}
&&\frac{\d}{\d t}d_{x}=\frac{d_{p}}{m}, 
\nonumber\\
&&\frac{\d}{\d t}d_{p}=-\frac{\gamma}{m}d_{p} + \eta (t)
-V'(d_{x})+\sum_{n=1}^{\infty}\frac{V^{(2n+1)} (d_{x})}{(2n)!!}
~d_{xx}^n,
\nonumber\\
&&\frac{\d}{\d t}d_{xx}=\frac{2}{m}d_{xp},
\nonumber\\
&&\frac{\d}{\d t}d_{pp}=-\frac{2\gamma}{m}d_{pp} 
-2\sum_{n=1}^{\infty}\frac{V^{(2n)} (d_{x})}{(2n-2)!!}
~d_{xx}^{n-1}d_{xp},
\nonumber\\
&&\frac{\d}{\d t}d_{xp}
=-\frac{\gamma}{m}d_{xp}+\frac{1}{m}d_{pp}
-\sum_{n=1}^{\infty}\frac{V^{(2n)} (d_{x})}{(2n-2)!!}
~d_{xx}^{n}.
\end{eqnarray}
This set of equations can be considerably simplified if 
the following change of variables will be made:
\begin{eqnarray}
\label{shaman1}
&&d_{x}=X,
\\
&&d_{p}=P,
\\
&&d_{xx}=Q^2,
\\
&&d_{pp}=\Pi ^2+\frac{\hbar ^2\sigma ^2}{4Q^2},
\\
&&d_{xp}=Q\Pi,
\end{eqnarray}
where $\sigma $ is chosen such that 
\BEQ
\label{shaman2}
\frac{1}{\sigma } = {\rm tr}(r^2)
\EEQ
With this change of variables Eqs.~(\ref{nippon}) will read
\begin{eqnarray}
\label{nippon-ke1}
&&\frac{\d}{\d t}X=\frac{P}{m}, 
\\ \label{nippon-ke2}
&&\frac{\d}{\d t}P=-\frac{\gamma}{m}P + \eta (t)
-\frac{\partial {\cal H} }{\partial X},
\\ \label{nippon-ke3}
&&\frac{\d}{\d t}Q=\frac{\Pi}{m},
\\ \label{nippon-ke4}
&&\frac{\d}{\d t}\Pi =-\frac{\gamma}{m}\Pi 
-\frac{\partial {\cal H} }{\partial Q},
\\
&&\frac{\d}{\d t}\sigma =-\frac{\gamma}{m}\sigma \,
\Rightarrow\, \sigma (t)=e^{-\gamma t/m}\sigma (0)=e^{-\gamma t/m},
\label{kosh}
\end{eqnarray}
where the initial state was chosen to be pure for simplicity: $\sigma
(0)=1$, and where ${\cal H}$ is an effective Hamiltonian:
\BEQ
\label{EH}
{\cal H}(P,X,\Pi,Q,t)=\frac{P^2}{2m}+\frac{\Pi ^2}{2m} +V(X)+
\sum_{n=1}^{\infty}\frac{V^{(2n)} (X)}{(2n)!!}
~Q^{2n}+\frac{\hbar ^2\sigma (t) ^2}{8mQ^2}.
\EEQ
The true Wigner function for the original quantum particle will read:
\BEA
W\left (x,p\right )=\int \d X~\d Q~\d P~\d \Pi ~
w\left (x,p\,|X,Q,P,\Pi
\right ){\cal P}(X,Q,P,\Pi),
\EEA
where ${\cal P}(X,Q,P,\Pi)$ is the ordinary probability distribution
of the classical random variables $X,Q,P$ and $\Pi$.

The physical meaning of this approach is now clear. The Hamiltonian
(\ref{EH}) corresponds to two classical particles with coordinates
$X,Q$ and momenta $P,\Pi$. Eqs.~(\ref{nippon-ke1}-\ref{nippon-ke4})
show that $X$-particle couples to the quantum bath through the damping
$\gamma P/m$ and noise $\eta (t)$. Although $\eta (t)$ is not an operator,
its correlator is still given by the quantum spectrum $K(t)$. 
$Q$-particle interacts with a classical bath at zero temperature, since
only
in this case a classical particle is subjected to damping, but not to
noise. The effective Hamiltonian ${\cal H}$ non-trivially couples 
these two particles. It is also time-dependent, though this dependence
is quite simple. 

Already the general form of the effective Hamiltonian leads us to the
following important observation. As follows from the derivation of the 
result, the purely classical case corresponds to $\hbar \to 0$, 
$\Pi\to 0$, $q\to 0$. Without damping one will have $\sigma ={\rm const}$,
which just reflects unitary evolution, where $r ^2$
is conserved.
Since $Q$, $\Pi $ typically have order ${\cal O}(\hbar )$, all terms in 
Eq.~(\ref{EH}) should disappear in the classical limit. 
For all terms besides the
last one this disappearance is clear. This last term can make the
classical
limit non-unique or even singular. This phenomenon is well-known in 
quantum chaos \cite{shavo,berry}. Moreover, even fairly simple integrable
systems can display singularities in the classical limit \cite{tatarinov}.
Now we observe that this dangerous term 
$\hbar ^2\sigma (t)^2/(8mQ^2)$ disappears with the characteristic time
$\gamma /m$, thereby ensuring the relatively straightforward classical
limit
in a damped system. Notice in this context that usually it is only the
noise,
which is believed to facilitate the classical limit, 
providing a mechanism for decoherence\cite{zurek}.

Notice that the equality
\BEQ
\label{larri}
d_{xx}d_{pp} - d_{xp}^2=\frac{\hbar ^2\sigma ^2}{4},
\EEQ
when $\sigma$ given by Eq.~(\ref{kosh}) is less than one, 
does not indicate a breaking of the uncertainty relations, 
since $\hat{r} $ itself
is not a density matrix. The correct uncertainty relation will read as
\BEQ
\label{larri1}
(\langle\langle \x ^2\rangle \rangle _{\eta }
-\langle\langle \x \rangle \rangle _{\eta }^2)
(\langle\langle \p ^2\rangle \rangle _{\eta }
-\langle\langle \p \rangle \rangle _{\eta }^2)
-\frac{1}{4}\left (
\langle\langle (\x - \langle\langle \x \rangle \rangle _{\eta }) 
(\p-\langle\langle \p \rangle \rangle _{\eta }) + 
(\p-\langle\langle \p \rangle \rangle _{\eta })
(\x - \langle\langle \x \rangle \rangle _{\eta })
\rangle \rangle _{\eta }
\right )^2\ge \frac{\hbar ^2}{4},
\EEQ
where $\langle\langle ...\rangle \rangle _{\eta }$ is the complete
average,
namely the average by $r $ (indicated with 
$\langle ...\rangle $), and by classical noise (indicated with
$\langle ...\rangle _{\eta }$).

These observable averages can be expressed as 
\BEQ
\label{tarantul1}
\langle\langle \x ^2\rangle \rangle _{\eta }
-\langle\langle \x \rangle \rangle _{\eta }^2
=\langle  d_{xx}\rangle _{\eta }+\langle  d^2_{x}\rangle _{\eta }
-\langle  d_{x}\rangle _{\eta }^2
\EEQ
\BEQ
\langle\langle \p ^2\rangle \rangle _{\eta }
-\langle\langle \p \rangle \rangle _{\eta }^2
=\langle  d_{pp}\rangle _{\eta }+\langle  d^2_{p}\rangle _{\eta }
-\langle  d_{p}\rangle _{\eta }^2
\label{tarantul2}
\EEQ
\BEQ
\label{tarantul3}
\frac{1}{2} (
\langle\langle (\x - \langle\langle \x \rangle \rangle _{\eta }) 
(\p-\langle\langle \p \rangle \rangle _{\eta }) + 
(\p-\langle\langle \p \rangle \rangle _{\eta })
(\x - \langle\langle \x \rangle \rangle _{\eta })
\rangle \rangle _{\eta } )
=\langle  d_{xp}\rangle _{\eta }+\langle  d_{x}d_{p}\rangle _{\eta }
-\langle  d_{x}\rangle _{\eta }\langle  d_{p}\rangle _{\eta }
\EEQ
Recall the situation with the exactly solvable harmonic potential. 
Here $d_{xx}$, $d_{xp}$, $d_{pp}$ tend to zero in the long-time limit,
being decoupled from $d_x$, $d_p$. Then 
Eq.~(\ref{larri1}) is obviously satisfied.

Let us briefly mention another aspect of the proposed scheme, which
can be interesting on general grounds. Two important 
length-scales are associated with any quantum system
\BEA
\label{tuli1}
&& L_c = \sqrt{D_{xx}}, \\
&& L_q=\sqrt{\frac{\hbar ^2D_{xx}}{4(D_{pp}D_{xx}-D^2_{xp})}},
\label{tuli2}
\EEA
where $D_{xx}$, $D_{pp}$ are dispersions of the coordinate and momentum,
and $D_{xp}$ is the corresponding cross-correlation.
These lengths quantify the quantum ($L_q$) and classical ($L_c$) 
aspects of the system, since due to uncertainty relation: 
$L_c/L_q\ge 1$, and the classical limit corresponds to 
$L_c/L_q\gg 1$. 

The analogous lengths for our variational scheme read
\BEA
\label{tuli3}
&& \tilde{L}_c = \sqrt{d_{xx}}, \qquad
\tilde{L}_q=\sqrt{\frac{\hbar ^2d_{xx}}{4(d_{pp}d_{xx}-d^2_{xp})}},
\\
&& \tilde{L}_c/\tilde{L}_q \sim e^{-\gamma t/m}.
\label{tuli4}
\EEA
The lengths $\tilde{L}_c$, $\tilde{L}_q $ still characterize
classical and quantum effects, but the uncertainty relation does 
not apply to them, due the above remarks.
Ratio (\ref{tuli4}) can now be much smaller than one, namely the quantum 
effects can be overdominating. The noise is needed to recover the 
uncertainty relation and to limit the overspread of quantum effects.

To conclude this section we will notice that recently an attempt was
made to construct a mean-field theory for a quantum brownian
particle \cite{terra}. However, the authors did not start from the
correct statement of the problem, and were led to an incorrect result
that their open quantum system can be still described by a wave
function (i.e., a pure state). In our notations this will amount to
put $\sigma (t)=1$ for all $t>0$ which is clearly incorrect in the
light of
Eqs.~(\ref{kosh}, \ref{iastreb4}). Besides the technical aspect
it contradicts to the general qualitative statement which we
draw after Eq.~(\ref{kosh}).

\section{Fokker-Planck equation}

Eqs.~(\ref{nippon-ke1}-\ref{nippon-ke4}) are still fairly complicated 
non-linear equations. To study them especially at low temperatures
we shall employ methods recently developed by two of us \cite{AN}.

We are looking for an equation describing the common probability
distribution
\BEQ
\label{P}
{\cal P}(y_1,y_2,y_3,y_4,t)=\langle
\delta (y_1-P(t))\delta (y_2-X(t))\delta (y_3-\Pi (t))\delta (y_4-Q(t))
\rangle _{\eta}
\EEQ
Using Eqs.~(\ref{nippon-ke1}-\ref{nippon-ke4}) and direct differentiation
one will get
\begin{equation}
\label{11}
\frac{\partial {\cal P}}{\partial t} =
\sum_{k=1}^4\frac{\partial (v_k P)}{\partial y_k} - \frac{\partial }
{\partial y_1}
\langle \delta (P(t)-y_1)\delta (X(t)-y_2)
\delta (\Pi (t)-y_3)\delta (Q(t)-y_4)\eta (t)\rangle _{\eta },
\end{equation}
where
\begin{eqnarray}
\label{11.1}
&&v_1=\frac{\gamma}{m}y_1(t)+\partial _{y_1}{\cal H}(y_1,y_2,y_3,y_4),
\\
&&v_2=-\frac{y_1}{m},
\\
&&v_3=\frac{\gamma}{m}y_3(t)+\partial _{y_2}{\cal H}(y_1,y_2,y_3,y_4),
\\
&&v_4=-\frac{y_3}{m}.
\end{eqnarray}

Since the gaussian noise is distributed with a functional
\begin{equation}
\label{fun}
\Omega [\eta ]\sim 
\exp-\frac{1}{2}\int\d t\d s\, \eta(t)K^{-1}(t-s)\eta(s),
\end{equation}
one has

\begin{equation}
\label{12}
\eta (t)\Omega [\eta ]
=-\int \d \top K(\top -t)\frac{\delta \Omega [\eta ]}{\delta \eta (\top
)},
\end{equation}

Substituting this equation into Eq.~(\ref{11}) one obtains after
functional
integration by parts:
\begin{eqnarray}
\label{111}
\frac{\partial {\cal P}}{\partial t} =
\sum_{k=1}^4\frac{\partial (v_k {\cal P})}{\partial y_k} - \frac{\partial
}
{\partial y_1}
\left \langle \frac{\delta }{\delta \eta (\top )}\left \{
\delta (P(t)-y_1)\delta (X(t)-y_2)
\delta (\Pi (t)-y_3)\delta (Q(t)-y_4)
\right \}
\right \rangle _{\eta },
\end{eqnarray}
To calculate the functional derivatives entering this equation, we notice
with the direct variation of the equations of motion:
\begin{equation}
\label{13}
\left [\begin{array}{r}
\delta P(t)/\delta \eta (t^{\prime})\\
\delta X(t)/\delta \eta (t^{\prime})\\
\delta \Pi (t)/\delta \eta (t^{\prime})\\
\delta Q(t)/\delta \eta (t^{\prime})
\end{array}\right ]
=\theta (t-t^{\prime})
\left \{ \exp 
\int^t_{t^{\prime}}\d u 
A(u)\right \} _{+}
\left [\begin{array}{r}
1 \\
0 \\
0 \\
0
\end{array}\right ], 
\end{equation}
\begin{equation}
\label{13a}
A(x)=\left (\begin{array}{rrrr}
-\gamma /m & -\partial ^2_{XX} {\cal H}& 0 & -\partial ^2_{XQ} {\cal H}\\
1/m & 0 & 0 & 0 \\
0 & -\partial ^2_{XQ} {\cal H} & -\gamma /m & -\partial ^2_{XX} {\cal H}\\
0 & 0 & 1/m & 0  
\end{array}\right ),
\end{equation}
where $\{ ... \}_{+}$ means the chronological ordering.

Substituting the last expression to Eq.~(\ref{11}) one gets
\begin{equation}
\label{14}
\frac{\partial {\cal P}}{\partial t}=
\sum_{k=1}^4 \frac{\partial }{\partial y_k}\left \{ v_k P
+\frac{\partial }{\partial y_1} 
\langle \delta (P(t)-y_1)\delta (X(t)-y_2)
        \delta (\Pi (t)-y_3)\delta (Q(t)-y_4)
\Phi _{k1} \{ x(t) \}
\rangle \right \},
\end{equation}
where $\Phi$ is the following $4\times 4$ matrix
\begin{equation}
\label{144}
\Phi(\{ x(t) \})=\int_0^t \d \top K(\top )
\left \{ \exp  \int^t_{t-\top }\d u A(x(u)) \right \}_{+} ,
\end{equation}
and $\Phi_{k1}$ is the corresponding matrix element.
This result is still exact, but untractable, since it involves 
the functional $\Phi$ of the history $\{ X(\top )\}$,
$\{ Q(\top )\}$ for $\top \le t$. 
In the classical limit one gets for $\top >0$ the white noise 
$K(\top )\to 2\gamma T\delta(\top )$, 
$$
\Phi _{k1}=\gamma T\delta _{k1},
$$
thus reproducing the corresponding classical Fokker-Planck-Kramers
equation. 
A closed equation for ${\cal P}$ can be obtained also in the harmonic 
case, where $A$ does not depend on $X(u),Q(u)$.
These two exact realizations prompt the way to proceed in the nonlinear
case. 
Since $K(\top )$ exponentially decreases for 
$t>{\rm max}(h\beta , \Gamma ^{-1})$, and this time can
assumed to be small in the quasiclassical domain,
one can make a Taylor expansion of the exponent
in Eq.~(\ref{144}), and keep only the first term:
\BEA
\label{145}
\left \{ \exp  \int^t_{t-s }\d u A(x(u))\right \} _{+}  
\approx e^{ s A(x(t)) }
\EEA 
Due to the $\delta$-function in eq. (\ref{14}), we may then replace 
the fluctuating $x(t), q(t)$ by the sure variables $x,q$, after which
$\Phi$ 
is no longer a fluctuating quantity, and can be taken
outside the averaging in eq. (\ref{14}), thus bringing a closed
equation for $P$.  $\Phi$ will be calculated with help 
of the folowing formula
\BEA 
\left [e^{ s A(x) }\right ]_{k1}=
-\oint \frac{\d z}{2\pi i} e^{sz}
\left [
\frac{1}{A-z\times 1}
\right ]_{k1},
\EEA
where $1$ is the $4\times 4$ unit matrix.

Our step (\ref{145}) is still exact for the harmonic initial potential
$V$, 
while in the general case with a characteristic scale of 
anharmonicity $L$ a condition
\BEQ
\label{res}
\frac{\gamma L^2}{\hbar}\gg 1,
\EEQ
is to be satisfied, which restricts the correlation time $h\beta$. 
If this time is not small enough, one can notice that the linear 
part of $A$ already suppresses exponentially the large values of $s$, 
and make the same expansion in Eq.~(\ref{14}). This is valid when the
nonlinearity of the potential is small with respect to its linear part.
Notice a certain nonperturbative 
aspect of the result, since the expansion was made inside of the exponent.

The final result that we obtain is a diffusion-type equation 
for $P$ itself:
\begin{eqnarray}
\label{ko1}
&&\frac{\partial {\cal P}(P,X,\Pi,Q,t)}{\partial t}=
-\frac{P}{m}\frac{\partial \PP}{\partial X}+\frac{\partial}{\partial P}
\left (\left[\frac{\gamma}{m}P+\frac{\partial {\cal H}}{\partial X}\right]
\PP \right )
-\frac{\Pi}{m}\frac{\partial \PP}{\partial q}+\frac{\partial}{\partial \Pi
}
\left (\left[\frac{\gamma}{m}\Pi +\frac{\partial {\cal H}}{\partial
Q}\right ]
\PP \right )
\nonumber\\
&&+\gamma D_{PP}(X,Q,t)\frac{\partial ^2\PP}{\partial P^2}
+\frac{\partial ^2}{\partial P\partial X}\left [
D_{XP}(X,Q,t)\PP\right ]
+\gamma D_{\Pi P}(X,Q,t)\frac{\partial ^2\PP}{\partial P\partial \Pi}
+\frac{\partial ^2}{\partial P\partial Q}\left [D_{QP}(X,Q,t)\PP\right ],
\end{eqnarray}
where 
we have changed $y_1\mapsto P$, $y_2\mapsto X$, $y_3\mapsto \Pi$, 
$y_4\mapsto Q$, and
diffusion coefficients $D_{PP}, {D}_{XP}, D_{\Pi P}, 
D_{QP}$ are instantaneous 
functions of $X$, $Q$ and $t$, and no longer functionals of the history.
Since the analytic structure of the diffusion coefficients is somewhat 
involved, it will be explained gradually. All diffusion coefficients
converge to finite values for large times. This convergence 
is exponential, and has a characteristic frequency
\BEQ
\label{char-om}
{\rm min}_{1\le k\le 4} [{\rm Re}(\omega _k)],
\EEQ
where
\begin{equation} 
\label{om12=}
\omega _{1,2}=\frac{\gamma}{2m}\left (
1\pm \sqrt{1+\frac{4mb_1}{\gamma ^2}}\,\,
\right ),
\label{omega}
\end{equation}
\begin{equation} 
\label{om34=}
\omega _{3,4}=\frac{\gamma}{2m}\left (
1\pm \sqrt{1+\frac{4mb_2}{\gamma ^2}}\,\,
\right ),
\end{equation}
and where
\BEQ
\label{kuno}
b_{1,2} = -\frac{1}{2}\left [
\partial _{xx}{\cal H}+\partial _{qq}{\cal H}\mp 
\sqrt{(\partial _{xx}{\cal H}-\partial _{qq}{\cal H})^2+\frac{4}{m^2}
[\partial _{xq}{\cal H}]^2}
\right ].
\EEQ
It seen that in order to have ${\rm Re}\,\omega _{k}\ge 0$, which is
necessary
for convergence, one has to require the conditions of local stability
\begin{eqnarray}
\label{stab}
&&\partial _{XX}{\cal H}+\partial _{QQ}{\cal H}\ge 0 ,
\nonumber \\
&&\partial _{XX}{\cal H}~\partial _{XQ}{\cal H}
\ge [\partial _{XQ}{\cal H}]^2.
\end{eqnarray}
Hereafter they will be assumed to be satisfied.

Let us now present explicit formulas for
the stationary values of the diffusion coefficients:
\begin{eqnarray}
\label{15}
D_{PP}(X,Q)=\frac{1}{m(b_1-b_2)}
\int^{\infty}_0\frac{\d \omega}{\pi}\bar{K}(\omega )\omega ^2
\left [
\frac{b_1+\partial _{qq}{\cal H} }
{(\omega ^2+\omega _1^2)(\omega ^2+\omega _2^2)}-
\frac{ b_2+\partial _{qq} {\cal H}  }
{(\omega ^2+\omega _3^2)(\omega ^2+\omega _4^2)}\right ],
\end{eqnarray}

\begin{eqnarray}
\label{16}
D_{XP}(X,Q)=\frac{1}{m(b_1-b_2)}
\int^{\infty}_0\frac{\d \omega}{\pi}\bar{K}(\omega )
\left [
\frac{(b_1+\partial _{qq}{\cal H}) (\omega ^2+b_1/m ) }
{(\omega ^2+\omega _1^2)(\omega ^2+\omega _2^2)}-
\frac{( b_2+\partial _{qq} {\cal H})(\omega ^2+b_2/m )  }
{(\omega ^2+\omega _3^2)(\omega ^2+\omega _4^2)}\right ],
\end{eqnarray}

\begin{eqnarray}
\label{17}
D_{\Pi P}(X,Q)=\frac{\partial _{xq}{\cal H}}{m(b_1-b_2)}
\int^{\infty}_0\frac{\d \omega}{\pi}\bar{K}(\omega )\omega ^2
\left [
\frac{1 }
{(\omega ^2+\omega _1^2)(\omega ^2+\omega _2^2)}-
\frac{1 }
{(\omega ^2+\omega _3^2)(\omega ^2+\omega _4^2)}\right ],
\end{eqnarray}

\begin{eqnarray}
\label{18}
D_{PQ}(X,Q)=\frac{\partial _{xq}{\cal H}}{m(b_1-b_2)}
\int^{\infty}_0\frac{\d \omega}{\pi}\bar{K}(\omega )
\left [
\frac{ \omega ^2 +b_1/m}
{(\omega ^2+\omega _1^2)(\omega ^2+\omega _2^2)}-
\frac{ \omega ^2 +b_2/m}
{(\omega ^2+\omega _3^2)(\omega ^2+\omega _4^2)}\right ],
\end{eqnarray}
where $\bar{K}(\omega )$ is the spectrum of $K(t)$, 
\begin{equation}
\label{2a2}
\bar{K}(\omega )= \hbar \gamma
\,\omega \coth \left (\frac{\hbar\omega\beta}{2}\right )
\frac{1}{1+(\omega /\Gamma )^2}.
\end{equation}
Let us indicate that 
the diffusion process (\ref{ko1}) is non-markovian.
In the purely classical limit the
time-dependence in the diffusion coefficients disappears, and
all diffusion coefficients besides $D_{PP}\to T$ disappear as well.
Only then Eq.~(\ref{ko1}) describes a markovian process.

\section{Reduced description}
Since Eq.~(\ref{ko1}) is still rather complicated, it is reasonable to
look for relatively simple limits. One of them is the overdamped limit,
which is characterized by large values of $\gamma $. 
In this case Eq.~(\ref{ko1}) can be reduced to an equation, which
describes
only slow variables $X$, $Q$. To proceed with this limit we shall define
the following moments
\BEQ
\label{tarzan}
\MM_{kl}(X,Q,t) = \int \d P~\d \Pi~P^k\Pi ^l\PP (X,P,Q,\Pi ,t),
\EEQ
and construct an equation for them starting from Eq.~(\ref{ko1}).
\begin{eqnarray}
\label{tarzan0}
\dot{\MM}_{kl}=&&-\frac{1}{m}\partial _X\MM_{k+1,l}
-\frac{1}{m}\partial _Q\MM_{k,l+1}
-\frac{(k+l)\gamma}{m} \MM_{k,l}-k\MM_{k-1,l}\,\partial _X{\cal H}
-l\MM_{k,l-1}\,\partial _Q{\cal H}\nonumber\\
&&+\gamma k(k-1)D_{PP}\MM_{k-2,l}-
k\partial _X[D_{XP}\MM_{k-1,l}]-l\partial _Q[D_{QP}\MM_{k,l-1}]
+\gamma kl\,D_{\Pi P}\MM_{k-1,l-1}.
\end{eqnarray}
Let us write down few first members of this hierarchy:
\BEQ
\label{tarzan1}
\dot{\MM}_{00}=-\frac{1}{m}\partial _X\MM_{10}-\frac{1}{m}\partial
_Q\MM_{01},
\EEQ
\begin{eqnarray}
\label{tarzan2}
\dot{\MM}_{10}=-\frac{1}{m}\partial _X\MM_{20}-\frac{1}{m}\partial
_Q\MM_{11}
-\frac{\gamma}{m} \MM_{10}-\MM_{00}\partial _X{\cal H}
-\partial _X[D_{XP}M_{00}],
\end{eqnarray}
\begin{eqnarray}
\label{tarzan3}
\dot{\MM}_{01}=-\frac{1}{m}\partial _X\MM_{11}-\frac{1}{m}\partial
_Q\MM_{02}
-\frac{\gamma}{m} \MM_{01}-\MM_{00}\partial _Q{\cal H}
-\partial _Q[D_{QP}\MM_{00}],
\end{eqnarray}
\begin{eqnarray}
\label{tarzan4}
\dot{\MM}_{20}=-\frac{1}{m}\partial _X\MM_{30}-\frac{1}{m}\partial
_Q\MM_{21}
-\frac{2\gamma}{m} \MM_{20}-2\MM_{10}\partial _X{\cal H}+2\gamma
D_{PP}\MM_{00}
-2\partial _X[D_{XP}\MM_{10}],
\end{eqnarray}
\begin{eqnarray}
\label{tarzan5}
\dot{\MM}_{02}=-\frac{1}{m}\partial _X\MM_{12}-\frac{1}{m}\partial
_Q\MM_{03}
-\frac{2\gamma}{m} \MM_{02}-2\MM_{01}\partial _Q{\cal H}
-2\partial _Q[D_{QP}\MM_{01}]
\end{eqnarray}
\begin{eqnarray}
\label{tarzan6}
&&\dot{\MM}_{11}=-\frac{1}{m}\partial _X\MM_{21}-\frac{1}{m}\partial
_Q\MM_{12}
-\frac{2\gamma}{m} \MM_{11}-\MM_{01}\partial _X{\cal H}
-\MM_{10}\partial _Q{\cal H}+\gamma D_{\Pi P}\MM_{00}
-\partial _X[D_{XP}\MM_{01}]-\partial _Q[D_{QP}\MM_{10}].
\nonumber\\ &&
\end{eqnarray}
In the first order of large $\gamma $ one can skip the time-derivatives in 
Eqs.~(\ref{tarzan1}, \ref{tarzan2}), since they have al least order 
${\cal O}(\gamma ^{-2})$. Further from Eqs.~(\ref{tarzan3}-\ref{tarzan3})
one gets the following approximate relations
\BEQ
\label{loki1}
\MM_{20}=mD_{PP}\MM_{00},
\EEQ
\BEQ
\label{loki2}
\MM_{02}={\cal O}(\gamma ^{-2}),
\EEQ
\BEQ
\label{loki3}
\MM_{11}=\frac{1}{2}mD_{\Pi P}\MM_{00}.
\EEQ
Notice from Eq.~(\ref{17}) that $D_{\Pi P}$ is of 
order $1/\gamma$ for large $\gamma$.
These equations are substituted in Eqs.~(\ref{tarzan1}, \ref{tarzan2}),
which
in combination with Eq.~(\ref{tarzan1}) brings the following reduced
equation
for the $\MM_{00}={\cal F}(X,Q,t)$, 
which is the probability distribution of the 
slow variables:
\BEA
\label{kolin}
&&\gamma\,\partial _t {\cal F} (X,Q,t)  =\partial _X
[{\cal F} \partial _X {\cal H} ]+\partial _Q
[{\cal F} \partial _Q {\cal H}]
+\partial _{XQ} [{\cal F} D_{\Pi P} ] +\partial _{XX}[ {\cal F}D_{XX}]
+\partial _{QQ}[{\cal F}D_{QP}],\\
&&
\label{kolin1}
D_{XX}=D_{XP}+D_{PP}.
\EEA
The first two terms in the r.h.s. of Eq.~(\ref{kolin}) are due to drift,
whereas other terms are responsible for the diffusion.
Let us now discuss this situation in details. In the classical limit,
which
is realized for sufficiently large temperatures or for $\hbar \to 0$,
one has $D_{XP}\to 0$, $D_{QP}\to 0$, $D_{\Pi P}\to 0$ and $D_{PP}\to T$.
Thus, Eq.~(\ref{kolin}) goes to the corresponding classical Fokker-Planck
equation. As Eqs.~(\ref{nippon-ke4}, \ref{kosh}, \ref{kolin}) show,
no noise is acting on the $Q$-particle, 
therefore in the classical case 
it just relaxes to zero and does not fluctuate at all. Therefore,
despite non-linearity of the potential $V(x)$, the 
classical variables $X,P$ decouples from the quantum variables $Q,\Pi$ and
tend to the classical Gibbs distribution. The quantum variables
disappear, as seen also from Eqs.~(\ref{loki2}, \ref{loki3}). 
Recall that the very fact of this homogeneous disappearance 
is connected with the exponential damping
(\ref{kosh}) of the singular term
$\hbar ^2\sigma (t) ^2/(8mQ^2)$
in the effective Hamiltonian (\ref{EH}).

On the other hand, in the quantum case Eq.~(\ref{kolin}) shows that
both variables $X$ and $Q$ become correlated and involved
in a common dynamics. Moreover, as follows from Eq.~(\ref{loki3}), there
is
a well-defined correlation between $\Pi $ and $P$. In particular, through
interaction with the classical variables quantum variables become coupled
to the thermal bath.

\section{Conclusion}

This paper was devoted to the mean-field (variational) theory of quantum
brownian motion. 
Mean-field methods are widely applied in quantum theory
\cite{heller,balian,shavo} and have an established range of validity.
Their general property is reduction of an initially 
quantum problem to an approximation 
involving only effective (mean-field) variables
with classical (commuting) behavior \cite{bobo}. The original 
quantum character of the problem is then reflected through an effective 
Hamiltonian. 
These properties of the mean-field description for closed systems 
were established also for more general cases 
(e.g. the time-dependent Hartree-Fock approximation), 
where the generated effective classical dynamics does not have the
canonical
form, but instead can be embodied into a more general Poisson structure
\cite{bobo}.

Being motivated by the effectiveness of the mean-field 
approach, 
we considered here its application to the problem of quantum brownian
motion,
which is the main representative model of quantum open systems. 

Our first step was to 
substitute the original quantum problem by an auxiliary semi-quantum one,
where only a part of degrees of freedom is quantum. In this exact step it
is 
possible to replace the operator-valued quantum noise by an auxiliary
classical gaussian noise, which has the same spectrum as the original
quantum noise. In that way we obtain the generalized Wigner-Moyal 
equation (\ref{moyal},
\ref{abdurahman}). In this step our description uses negative
probabilities
in the sense that the unaveraged density matrix (\ref{iastreb2}) 
does have negative eigenvalues.
However, no unphysical results appear in the level of observables
quantities. The negative probabilities appear as the cost for having 
explicitly classical elements in a quantum dynamics.
This situation is reminiscent of the ordinary Wigner function which also
cannot be interpreted as a probability density, but which shares some of
its 
properties and does predict correct quantum mechanical averages. 

At the second step
of our description we applied the Gaussian approximation to the unaveraged
Wigner function. By this procedure the initial quantum stochastic problem 
became reduced to a problem of two classical particles with friction and 
classical noise.
This noise is nevertheless not white, but is
correlated with the same spectrum as the original quantum noise. Further 
investigation allowed us to uncover an important role played by friction
in establishing classical aspects of the problem. It appeared that a
singular
$\hbar$-dependent term in the effective Hamiltonian (\ref{EH}) is
diminished 
by friction with a characteristic time (\ref{kosh}) inversely 
proportional to the damping 
coefficient. This ensures the existence of the unique
and homogeneous classical limit for times larger than the above
characteristic time. This fact is contrasting with the non-commutation
of the classical ($\hbar\to 0$) and long-time $t\to\infty$ limit for 
(closed) Hamiltonian systems \cite{berry}.
At low temperatures, where the quantum effects are essential, the dynamics
of the mean-field degrees of freedom is essentially different, because
they
become correlated with each other. This is shown in particular by 
Eqs.~(\ref{loki3}, \ref{kolin}). In other words, the effective classical
dynamics still contains $\hbar/T$ as a correlation time of the classical
noise, and therefore displays essential different dynamical behavior for
high and low temperatures \cite{AN}.

In the present paper we restricted ourselves to the general framework
of the mean-field quantum Brownian motion. More specific applications,
e.g. for open many-body systems, are planned to consider in future.
Finally, it is hoped that the paper will open a road for applications
of mean-field methods in quantum dissipative systems.

\acknowledgments 
A.E. A. acknowledges support by NATO.


\begin{thebibliography}{99}
\vspace{-1cm}

\bibitem{ballentine}L.E. Ballentine, {\it Quantum Mechanics}, 
Prentice-Hall, 1990.

\bibitem{berry} M.V. Berry, in {\it Chaos and Quantum Physics},
Proceedings of the Les Houches Summer School 1989, ed. by
M.J. Jianonni, A. Voros, J. Zinn-Justin, North Holland,
Amsterdam, 1991.

\bibitem{tatarinov}V.I. Tatarskii, Sov. Phys. Usp., {\bf 26} 311 (1983)

\bibitem{heller} E. Heller, in {\it Chaos and Quantum Physics},
Proceedings of the Les Houches Summer School 1989, ed. by
M.J. Jianonni, A. Voros, J. Zinn-Justin, North Holland,
Amsterdam, 1991.

\bibitem{balian}R. Balian and M. Veneroni, Ann. Phys., {\bf 187} 29 (1988)

\bibitem{shavo} A.K. Pattayanak and W.C. Schieve, Phys. Rev. Lett.,
{\bf 72} 2855 (1994). Phys. Rev. E, {\bf 50} 3601 (1994). 

\bibitem{klaud} P. Kramer, M. Saraceno, {\it Geometry of Time-Dependent 
variational Principle in Quantum Mechanics}, (Springer-Verlag, Berlin,
1981)

\bibitem{weiss} U. Weiss, {\it Quantum Dissipative Systems}, 
(World  Scientific, Singapore, 1993).

\bibitem{gardiner} C. Gardiner, {\it Quantum Noise}, 
(Springer, Berlin, 1991).


\bibitem{sen}
G.W. Ford, M. Kac and P. Mazur, J. Math. Phys.
{\bf 6} 504 (1965).

\bibitem{muck}W. Muckenheim, Phys. Rep. {\bf 133} 337 (1986)

\bibitem{haake}
F. Haake and R. Reibold, Phys. Rev. A {\bf 32} 2462 (1985).
(1981).

\bibitem{zurek}
J.P. Paz, S. Habib, W.H. Zurek, Phys. Rev. D, {\bf 47}, 488, (1993).

\bibitem{AN}
A.E. Allahverdyan and Th.M. Nieuwenhuizen, 
Phys. Rev. Lett {\bf 85}, 1799 (2000); cond-mat/0011389

\bibitem{terra} W.V. Liu and W.C. Schieve, Phys. Rev. Lett., {\bf 78}
(1997) 3278; chao-dyn/9703012.

\bibitem{grad}H. Grad, {\it Principles of the Kinetic Theory of Gases},
in {\it Handbuch der Physik}, ed. by S. Fl\"{u}gge, Springer-Verlag (1958)

\bibitem{bobo}R. Balian and M. Veneroni, Ann. Phys., {\bf 195} 324 (1988)


\end{thebibliography}
\end{document}